# Anomalous Nonlinear Magnetoconductivity in van der Waals Magnet CrSBr

*Junhyeon Jo\*, Manuel Suárez-Rodríguez, Samuel Mañas-Valero, Eugenio Coronado, Ivo Souza, Fernando de Juan, Fèlix Casanova, Marco Gobbi\*, and Luis E. Hueso\**

J. Jo, M Suárez-Rodríguez, F. Casanova, L. E. Hueso
CIC nanoGUNE BRTA, 20018 Donostia-San Sebastián, Spain
E-mail: j.jo@nanogune.eu; l.hueso@nanogune.eu

S. Mañas-Valero, E. Coronado
Instituto de Ciencia Molecular (ICMol), Universitat de València, 46980 Paterna, Spain

F. de Juan
Donostia International Physics Center (DIPC), 20018 Donostia-San Sebastián, Spain

I. Souza, M. Gobbi
Centro de Física de Materiales (CFM-MPC) Centro Mixto CSIC-UPV/EHU, Donostia-San Sebastián, Spain
E-mail: marco.gobbi@ehu.eus

I. Souza, F. de Juan, F. Casanova, M. Gobbi, L. E. Hueso
IKERBASQUE, Basque Foundation for Science, 48009 Bilbao, Spain

## Abstract

Nonlinear magnetoconductivity (NLMC) is a nonreciprocal transport response arising in non-centrosymmetric materials. However, this ordinary NLMC signal vanishes at zero magnetic field, limiting its potential for applications. Here, we report the observation of an anomalous NLMC controlled by internal order parameters such as the magnetization or Néel vectors. We achieve this response by breaking both inversion and time-reversal symmetry in artificial van der Waals heterostructures based on the magnetic CrSBr and insulating hBN. The nonreciprocal signal can be tuned between two different states in ferromagnetic monolayer CrSBr and among four different states in antiferromagnetic bilayer CrSBr, thanks to its metamagnetic transition. Remarkably, this output signal in the ferromagnetic (antiferromagnetic) state of CrSBr is three (one) orders of magnitude higher than those previously measured. A conductivity scaling analysis reveals the Berry connection polarizability as the origin of the anomalous NLMC. Our results pave the way for high-frequency rectifiers with magnetically switchable output polarity as well as for an efficient electrical readout of the magnetic state of antiferromagnetic materials.

# 1. Introduction

Electrical response of materials is inherently tied to their symmetries. For example, breaking time-reversal symmetry ($\mathcal{T}$) through the application of a magnetic field $\boldsymbol{B}$ in conductive materials causes the emergence of a voltage transverse to a current direction: this is the well-known ordinary Hall effect.[1] In ferromagnets (FM), as well as in antiferromagnets (AFM) where the internal magnetic order breaks $\mathcal{T}$, the anomalous Hall effect emerges with a transverse voltage persisting even at zero magnetic field.[2] Recent studies have shown that breaking inversion symmetry ($\mathcal{P}$) leads to nonreciprocal transport effects.[3-8] One such feature is nonlinear magnetoconductivity (NLMC),[3,5] also known as unidirectional magnetoresistance, which arises in noncentrosymmetric materials when an applied magnetic field breaks $\mathcal{T}$.[9-19] The NLMC causes the resistance $R$ to depend on the current direction under an applied magnetic field, generating a second-order longitudinal resistance $R^{(2)}$ that changes its sign when the magnetic field direction is reversed.

Just as the anomalous Hall effect is the magnetization-driven counterpart to the ordinary Hall effect, an analogous version of NLMC, driven by an internal magnetic order, can be expected in magnetic materials.[20] However, the stringent crystal symmetry constrains has limited the observation of this effect to CuMnAs[21] and $MnBi_2Te_4$,[22-25] which require delicate epitaxial growth. It is hence essential to find a broader range of magnetic systems in which the anomalous NLMC can be explored and exploited. This objective can be achieved by engineering two-dimensional van der Waals heterostructures with on-demand broken symmetry.

In this regard, CrSBr is a van der Waals magnet that provides an ideal platform to explore the interplay between broken structural inversion symmetry, magnetism, and nonreciprocal responses.[26] In its monolayer form, CrSBr exhibits FM order, whereas bilayer CrSBr shows AFM order.[26-28] Interestingly, the bilayer displays a metamagnetic transition that switches the material from an AFM state at low fields to a FM state at high fields. CrSBr has a centrosymmetric crystal structure. However, the magnetic configuration in the AFM order breaks $\mathcal{P}$ of CrSBr,[26] while the FM state of CrSBr preserves its $\mathcal{P}$. The reduced dimensionality of ultrathin two-dimensional materials enables precise engineering of their symmetry,[8,29] potentially leading to breaking $\mathcal{P}$ in the FM state of CrSBr through the formation of tailored interfaces. This would make CrSBr a versatile system for probing the anomalous NLMC effect in both FM and AFM states.

Here, we report the observation of a large and layer-dependent anomalous NLMC in the van der Waals magnet CrSBr. The effect is tunable through the manipulation of both the magnetization vector ***M*** and Néel vector ***N***. Specifically, monolayer CrSBr exhibits a hysteretic behavior in $R^{(2)}$, displaying a difference for opposite magnetization directions at zero magnetic field. More interestingly, bilayer CrSBr displays unprecedented four different $R^{(2)}$ states. At zero magnetic field, two distinct states are observed, each corresponding to opposite orientations of the Néel vector. Additionally, at magnetic fields exceeding the metamagnetic transition, two additional states emerge, corresponding to the two FM states with opposite magnetization directions. The recorded output value in the FM (AFM) state is three (one) orders of magnitude higher than those found in magnetic topological insulators.[22,23,25,30] A conductivity scaling analysis reveals a scattering-independent behavior in our systems, indicating that the Berry connection polarizability[25,31-33] is responsible for the recorded signals. This anomalous NLMC enables switchable $R^{(2)}$ signals at zero magnetic field, representing a potential significant advance for applications such as reconfigurable high-frequency rectification for energy harvesting,[34-38] and electrical readout of antiferromagnetic memories.[21]

## 2. Results
### 2.1. Nonlinear magnetoconductivity under different magnetic orders

The NLMC effect was initially reported for nonmagnetic-noncentrosymmetric systems.[3,4] The resistance in these systems, in addition to the conventional magnetoresistance which scales with the square of an applied magnetic field ($B^2$), depends on the product of a magnetic field and a current *I*, and it is described by the equation $R = R_0(1 + \alpha B^2 + \gamma IB)$, where $R_0$ is the resistance at zero field and $\alpha$ and $\gamma$ indicate the coefficients in each term, respectively.[5] Consequently, the second-order resistance, $R^{(2)} = [R(+I) - R(-I)]/2 = R_0 \gamma IB$, increases linearly with both the magnetic field and the current. This expression is a particular case of a more general tensorial relation applicable to all systems with broken $\mathcal{P}$, in which the second-order resistance $R^{(2)}$ to linear order in ***B*** is given by $R^{(2)}_{ij} = R^{(2)B}_{ijkl} I_k B_l$ (see Note S1 for details), which represents the fact that NLMC can occur for different orientations of magnetic field and current directions. For instance, in a chiral system,[3,18,19,39-45] the NLMC, in particular known as electrical magnetochiral anisotropy, is typically strong when the field is parallel to the current (**Figure 1a**), which corresponds to $R^{(2)} = R_0 \gamma_\parallel IB$ with $\gamma_\parallel \equiv R^{(2)B}_{iiii}/R^{(1)}_{ii}$ and $R_0 \equiv R^{(1)}_{ii}$, while in this work we focus on NLMC where the field is perpendicular to the current, $R^{(2)} = R_0 \gamma_\perp IB$ with $\gamma_\perp \equiv R_{iiij}^{(2)B}/R_{ii}^{(1)}$ with $i \neq j$.

In this work, CrSBr is employed to study the anomalous version of NLMC, where $\mathcal{T}$ is broken by the internal magnetic order rather than by an external magnetic field. In FM monolayer CrSBr (Figure 1b), the magnetization-induced nonreciprocal response is $R^{(2)} = R_0 \gamma_M IM$, where $M$ is the magnetization vector that replaces $B$ and, in analogy to $\gamma_\perp$, $\gamma_M$ is the anomalous NLMC coefficient for the FM state when the longitudinal resistance is measured along the *a*-axis and the magnetization vector is aligned with the *b*-axis. The resistance therefore follows the simplified equation $R = R_0(1 + \gamma_M IM)$. Experimentally, $\gamma_M$ is allowed by breaking $\mathcal{P}$ due to the presence of a top hexagonal boron nitride (hBN) layer (see details in Note S1). Notably, two $R^{(2)}$ states with opposite signs are expected for opposite directions of $M$ even at zero magnetic field, as sketched in Figure 1b.

Bilayer CrSBr presents an AFM configuration which intrinsically breaks $\mathcal{P}$ (Figure 1c),[26] leading naturally to the emergence of the second-order response (see details in Note S1). In this case, the nonreciprocal resistance is $R^{(2)} = R_0 \gamma_N IN$, where $N$ is the Néel vector and $\gamma_N$ is the anomalous NLMC coefficient for the AFM state when the longitudinal resistance is measured along the *a*-axis and the Néel vector is aligned with the *b*-axis. The resistance therefore follows the simplified equation $R = R_0(1 + \gamma_N IN)$ at low magnetic field regions. At higher fields, the bilayer CrSBr undergoes a metamagnetic transition, entering a FM state. This state results in two additional resistance states, similar with the monolayer case. Therefore, the bilayer CrSBr is uniquely characterized by four magnetic-field-addressable $R^{(2)}$ states. This control of anomalous NLMC leveraging both magnetization and Néel vectors in a single system had been until now an unfulfilled challenge.

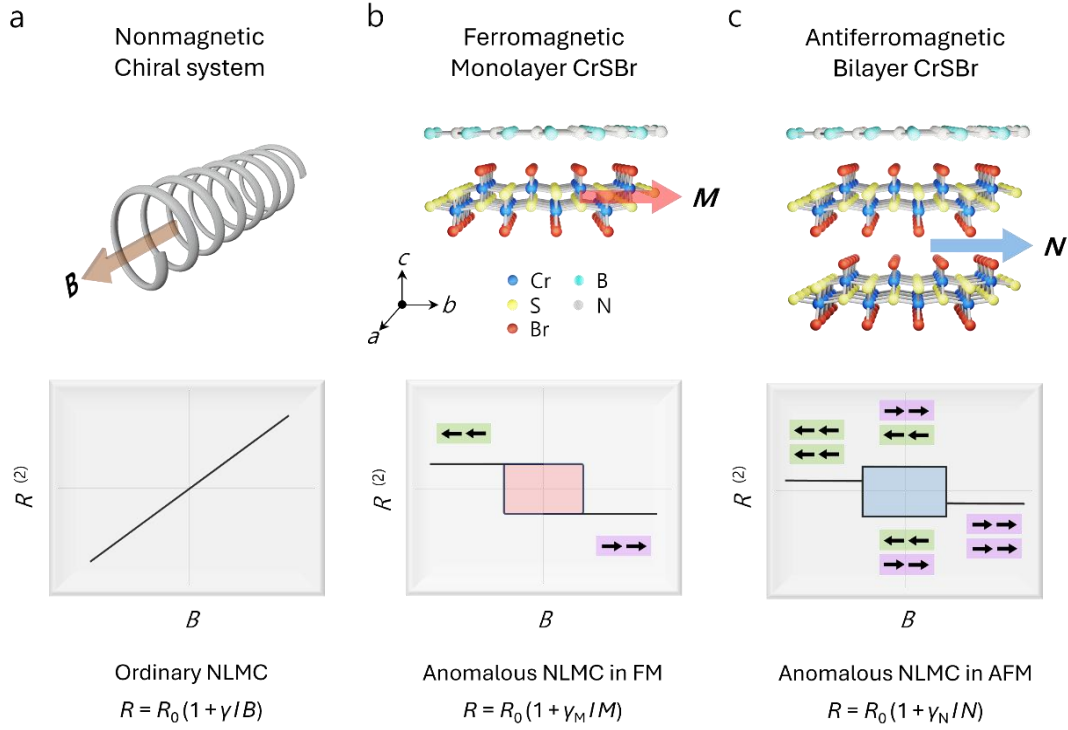

**Figure 1.** Nonlinear magnetoconductivity (NLMC) in systems with different magnetic response. a) Ordinary NLMC in a nonmagnetic, chiral material. Second-order longitudinal resistance is defined as $R^{(2)} = [R(+I) - R(-I)]/2$. This response is linear with the applied magnetic field $B$ ($= \mu_0 H$) and the electrical current ($I$), and it is described by the expression $R = R_0(1 + \gamma I B)$, where $R_0$ is a resistance at zero field and $\gamma$ is the coefficient of the nonreciprocal response. b) Anomalous NLMC response in ferromagnetic (FM) monolayer CrSBr. The presence of a top hexagonal boron nitride (hBN) layer breaks inversion symmetry of the centrosymmetric FM CrSBr, leading to an anomalous nonreciprocal response controlled by the magnetization vector $M$. c) Anomalous NLMC response in antiferromagnetic (AFM) bilayer CrSBr. In this case, the Néel vector $N$ governs the two states in the AFM region in which inversion symmetry is intrinsically broken. In addition, the magnetization vector gives rise to two additional states in the FM state similar with the monolayer case, for a magnetic field above the metamagnetic transition of the CrSBr. Similar analytic expressions describe the response in each case. In the magnetic systems, the anomalous response is no longer controlled by the external $B$ but by the internal vectors $M$ and $N$ for the FM and AFM phase, respectively.

## 2.2. Anomalous NLMC in ferromagnetic CrSBr(1L) system

**Figure 2** shows the NLMC response in the FM system, CrSBr(1L)/hBN. To perform electrical transport measurement, the CrSBr/hBN heterostructure was placed on top of prepatterned Au electrodes using a polymer-assisted dry transfer method (see details in Methods). As shown in the scanning electron microscope image of the CrSBr/hBN device (Figure 2a), a current was applied along the *a*-axis of a CrSBr flake, and a longitudinal voltage ($V$) was measured under an in-plane magnetic field. In this configuration, magnetoresistance (MR) was measured using a direct current (of both polarities, $\pm I$). A magnetic field was applied along the *b*-axis, the magnetic easy axis, of CrSBr[28,46,47] except where otherwise noted.

Figure 2b displays MR curves under the two electrical current polarities, +$I$ (top) and –$I$ (bottom), at 2 K. Distinctive hysteresis loops are observed with a coercive field of 0.1 T, reflecting the FM nature of CrSBr(1L). When the two direct current measurements are averaged as $R^{(1)} = [R(+I) + R(-I)]/2$ to extract the purely linear response, the hysteresis loop disappears (Figure S1).

The anomalous NLMC, represented by the second-order response $R^{(2)}$, shows an hysteretic behavior with two well-defined resistance states for positive and negative magnetic fields (Figure 2c). This anomalous nonreciprocal resistance state is deterministically controlled by the magnetization vector of the FM CrSBr(1L). Second-harmonic measurement of $R^{2\omega}$ with an alternating current and a lock-in amplifier exhibits an equivalent response (Figure S2). As expected, this effect only appears under broken $\mathcal{P}$ condition, which is created by the top hBN layer. MR results on an equivalent device, but without a top hBN layer, do not show any NLMC response (Figure S3).

A comparison of the magnetic field dependent measurements of $R^{(2)}$ along different crystallographic axis is plotted in Figure 2d, reflecting the relation between the magnetization and current directions. The black curve shows the data plotted in Figure 2c up to a higher magnetic field sweep along the *b*-axis, highlighting that the $R^{(2)}$ saturation value does not change. In contrast, the behavior of $R^{(2)}$ measured under a magnetic field sweep along the crystallographic *a*-axis (red) shows some distinctive features. It also displays a hysteretic behavior with its maximum amplitude at zero field, while the output decreases with magnetic field and disappears above 1 T. The features can be understood by considering that, at small magnetic fields the magnetic anisotropy aligns the magnetization with the *b*-axis, the easy axis of CrSBr. In this case, the anomalous NLMC is maximized since the magnetization is perpendicular to the current (***M*** ⊥ ***I***). However, at high fields, the magnetization is fully aligned with the *a*-axis, and the output vanishes since ***M*** ∥ ***I*** (see the full angular dependence in Figure S4). From here, we can extract two main conclusions: in the first place, that ***M*** and not ***B*** is the driving parameter of the nonlinear response; in the second place, we need ***M*** ⊥ ***I*** to obtain an anomalous NLMC in agreement with the constraints imposed by the $C_{2v}$ symmetry of our system (see Note S1).

Figure 2e shows $R^{(2)}$ at zero magnetic field as a function of the applied current and for each of the two orientations of the magnetization vector along the *b*-axis in CrSBr(1L)/hBN. The FM–I (FM–II) notation refers to the remanent state at zero magnetic field after the magnetic field is swept from a negative (positive) field region, corresponding to the magnetic configuration sketched in Figure 2c. The magnitude of the $R^{(2)}$ signal linearly increases with the applied current, but its sign is only determined by the magnetization state of the CrSBr. Thus, the CrSBr(1L)/hBN system presents anomalous NLMC in the FM state, which is governed by the relation $R = R_0(1 + \gamma_M IM)$. The coefficient of anomalous NLMC in the FM state, estimated by $\gamma_M M = R^{(2)}/(R_0 I)$. When we convert the coefficient for a current density ($j$), the normalized coefficient $\gamma'_M M = R^{(2)}/(R_0 j)$ is estimated as $5.93 \times 10^{-12}$ A$^{-1}$m$^2$. Remarkably, the output second-order resistance normalized by current density ($R^{(2)} / j$) is $4.06 \times 10^{-6}$ ($\Omega$m$^2$/A), three orders of magnitude higher than the result for a FM edge state of topological insulators (see the comparison in Table S1).[22,30]

The recorded anomalous $R^{(2)}$, which is defined as the maximum amplitude of the hysteresis loop at zero field, gradually decreases as temperature increases and disappears above 100 K (Figure 2f), close to the Curie temperature of CrSBr(1L).[27] In other words, the anomalous NLMC response is absent in the paramagnetic state of CrSBr(1L)/hBN, highlighting the requirement of the time-odd nature of CrSBr (see additional information for the Curie temperature in Figure S5).

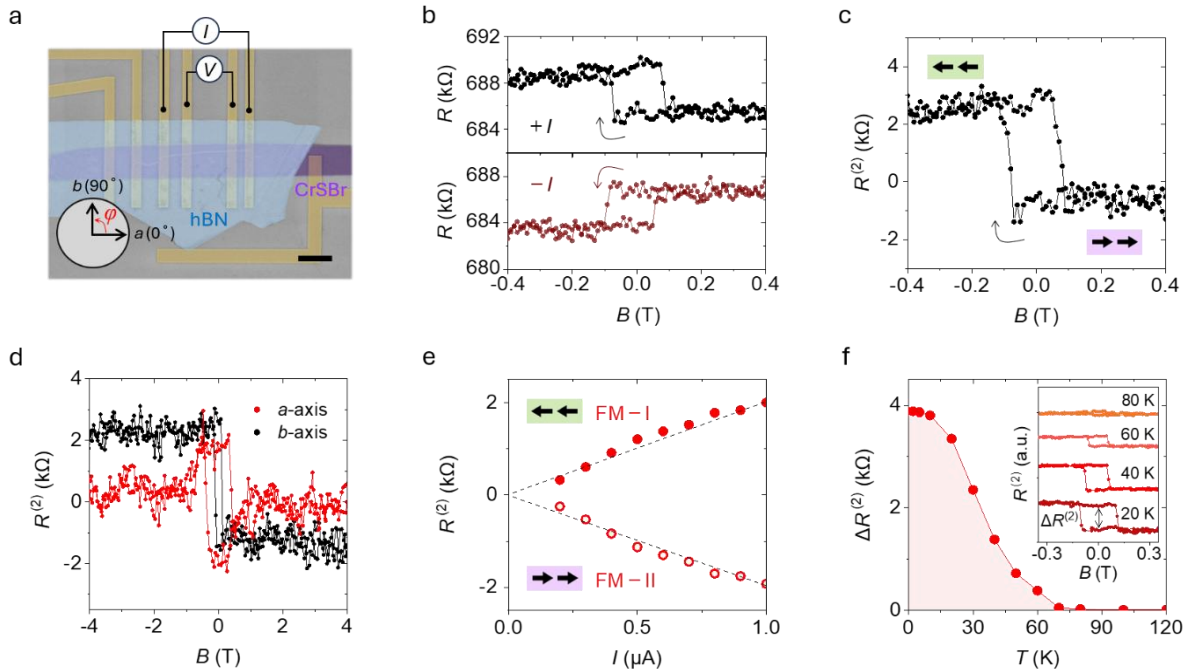

**Figure 2**. Anomalous nonlinear magnetoconductivity in the ferromagnetic system CrSBr(1L)/hBN. a) Scanning electron microscopy image of a CrSBr(1L)/hBN device. An electrical current ($I$) is applied along the $a$-axis of the CrSBr flake, while voltage ($V$) probes measure the longitudinal resistance. The scale bar is 2 μm. b) Resistance as a function of a magnetic field (*B*) applied along the crystallographic *b*-axis of CrSBr. The graphs present the measured resistance while applying a positive (top) or a negative (bottom) current of 1 μA at 2 K. Opposite hysteresis loops are observed for opposite current polarity. c) Second-order resistance, $R^{(2)} = [R(+I) - R(-I)]/2$, as a function of a magnetic field, obtained from the data in b. The presence of anomalous NLMC at zero magnetic field is clear. d) Second-order resistance as a function of a magnetic field along the crystallographic *a*-axis and *b*-axis of CrSBr. The $R^{(2)}$ reaches the maximum at zero magnetic field under the field sweep along the *a*-axis (red) since the magnetization naturally aligns to the magnetic easy axis of CrSBr (*b*-axis). This is the case when the magnetization is perpendicular to the current. When the magnetization is aligned with the *a*-axis above 1 T, the $R^{(2)}$ vanishes as the magnetization is parallel to the current. e) Current-dependent $R^{(2)}$ for both orientations of the magnetization vector in the FM state at zero magnetic field, denoted as FM−I and FM−II. The $R^{(2)}$ signal exhibits a linear dependence with the applied current, but its sign depends on the orientation of the magnetization. f) $\Delta R^{(2)}$ as a function of temperature. $\Delta R^{(2)}$ is defined as the maximum amplitude of the hysteresis loop at zero field, as shown in the inset.

## 2.3. Anomalous NLMC in antiferromagnetic CrSBr(2L) system

**Figure 3** presents the anomalous NLMC in the CrSBr(2L)/hBN. The MR curves for a positive and negative current show the metamagnetic transition from the AFM state into the FM state as the magnetic field increases (Figure 3a). When a positive current (top) is applied while sweeping the magnetic field to the negative direction (solid symbols), the resistance state changes from low to high near zero field, and goes back to the low resistance state at a field of –0.3 T. The curve shift from zero magnetic field during the metamagnetic transition could be attributed to spin torque or bias effect during the spin-flip transition, or substrate effects as reported in layered metamagnets.[46,48,49] In a similar way, sweeping the field to the positive direction shows the metamagnetic transition as well, but this time it occurs in the opposite field region (open symbols). On the other hand, when a negative current (bottom) is applied during a magnetic field sweep, resistance differences are detected compared to the data with the positive current, in each AFM and FM state.

$R^{(1)}$ shows a well-defined metamagnetic transition featuring the AFM and FM states of the CrSBr(2L) (Figure 3b), as previously reported in multilayer CrSBr.[46,47] By contrast, $R^{(2)}$ presents four distinct resistances states (see Figure 3c). In the AFM region for magnetic fields lower than 0.3 T, two AFM $R^{(2)}$ states exist, distinguishable by the field sweeping direction. A higher resistance state is achieved when the field is swept from –9 T, which we denote as AFM–I (present from 0 T to +0.3 T). A lower resistance state, AFM–II (present from 0 T to –0.3 T), is obtained during the field sweeping from +9 T. This deterministic orientation of Néel vector could be attributed to the asymmetric structure of the CrSBr device, sandwiched by different top (hBN) and bottom ($SiO_2$) dielectric layers. This structure might affect the magnetic anisotropy of CrSBr and lead to a preferred Néel vector orientation. The identical but opposite values of $R^{(2)}$ indicate that they arise from the opposite orientations of the Néel vector of CrSBr(2L). Accordingly, the anomalous NLMC allows to electrically read the magnetic state of an AFM material and to utilize the different orientation of Néel vector, which is typically a very complex task. In the FM state, for a magnetic field higher than 0.3 T in absolute value, two additional $R^{(2)}$ states are observed. These two states are denoted as FM–I (sweeping from the negative field region) and FM–II (sweeping from the positive field region). This hysteretic behavior is equivalent to the one previously observed in the CrSBr(1L)/hBN, as the top hBN layer breaks $\mathcal{P}$ of CrSBr(2L) and generates two FM resistance states depending on the orientation of the magnetization vector. Overall, we obtained four NLMC states in the CrSBr(2L)/hBN system.

The resultant $R^{(2)}$ values in both the AFM and FM states show as well the expected linear dependence on the applied current (Figure 3d,e). As noted above, the sign of $R^{(2)}$ is opposite depending on the orientations of $N$ and $M$, confirming the relations $R = R_0(1 + \gamma_N IN)$ for an AFM system and $R = R_0(1 + \gamma_M IM)$ for a FM system. The normalized coefficient of anomalous NLMC in the AFM state, estimated by $\gamma'_N N = R^{(2)}/(R_0\, j)$ is $4.12 \times 10^{-11}$ $A^{-1}m^2$, and the coefficient in the FM state, $\gamma'_M M = R^{(2)}/(R_0\, j)$ is $1.90 \times 10^{-11}$ $A^{-1}m^2$. The normalized ($R^{(2)}/j$) output signal in the AFM state is $1.41 \times 10^{-5}$ $\Omega m^2 A^{-1}$, which is one order of magnitude higher than the value reported for an AFM topological insulator,[23,25] while the obtained output in the FM state is $6.80 \times 10^{-6}$ $\Omega m^2 A^{-1}$ that is on the same order of the FM CrSBr(1L) case (see the comparison in Table S1).

These anomalous NLMC responses in the CrSBr(2L) persist up to 140 K, the Néel temperature of CrSBr (Figure 3f).[27] The disappearance of the nonreciprocal response in the paramagnetic phase of CrSBr(2L) again clearly demonstrates that the response is linked to the magnetic order of CrSBr.

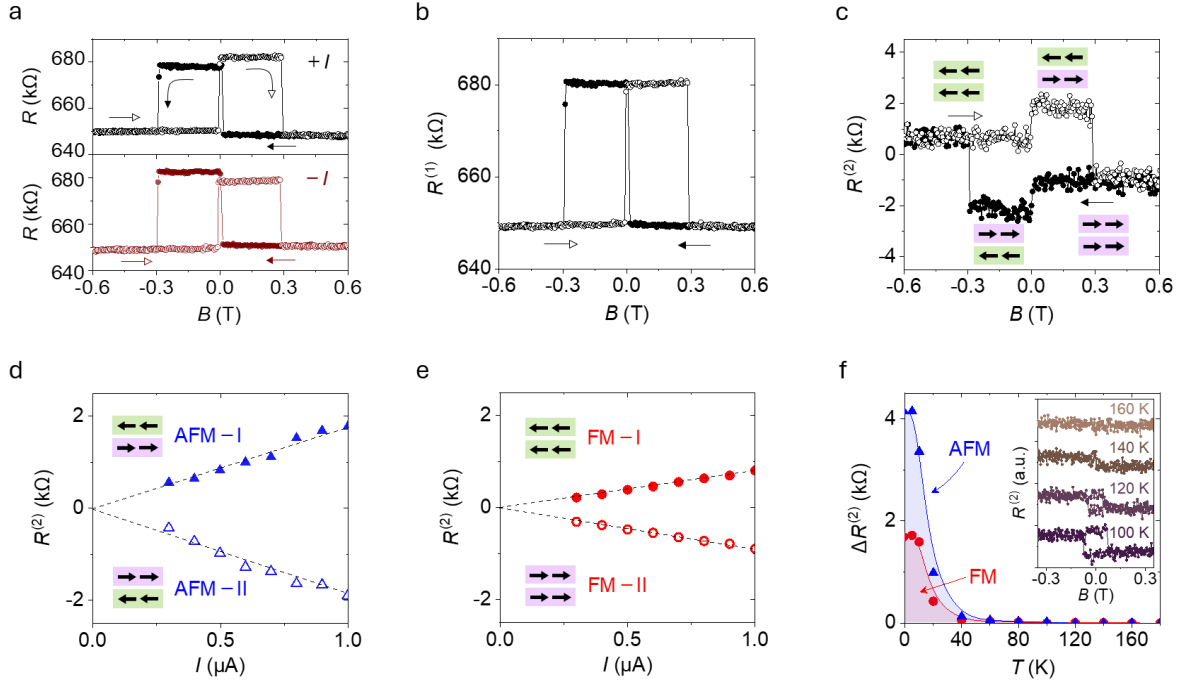

**Figure 3**. Anomalous nonlinear magnetoconductivity in both antiferromagnetic and ferromagnetic states of CrSBr(2L)/hBN. a) Resistance measured with a positive (+*I*) and a negative (−*I*) current of 1 μA along the *a*-axis at 2 K, and as a function of an applied magnetic field (*B*) parallel to the *b*-axis. Different electrical current polarity results in different resistance curves, *R*(+*I*) and *R*(−*I*). b) Average resistance, defined as $R^{(1)} = [R(+I) + R(-I)]/2$, as a function of the magnetic field, obtained from the data in a. Sharp magnetoresistance curves are observed, featuring the metamagnetic transition between the AFM and FM states of CrSBr(2L). c) Second-order resistance, measured as $R^{(2)} = [R(+I) - R(-I)]/2$, as a function of the magnetic field. In the AFM region, the opposite orientations of the Néel vector result in opposite $R^{(2)}$ values stemming from the intrinsic broken inversion symmetry of CrSBr(2L). In the FM state, the opposite orientations of the magnetization vector cause opposite $R^{(2)}$ values as inversion symmetry is broken by the top hBN layer. d,e) Current-dependent $R^{(2)}$ in both the AFM and FM states. The observed $R^{(2)}$ is linear with the applied current, but its sign changes with the opposite orientation of the Néel and magnetization vectors, respectively. f) Maximum amplitude of the hysteresis loop, $\Delta R^{(2)}$, as a function of temperature for both the AFM and FM states. The amplitude gradually decreases and disappears around 140 K, the Néel temperature of the AFM CrSBr(2L).

## 2.4. Microscopic origin of anomalous NLMC in CrSBr

Finally, we investigated the microscopic origin of the anomalous NLMC in CrSBr. A scaling relation between the second-order conductivity ($\sigma^{(2)}$) and the linear conductivity ($\sigma$) can elucidate which mechanisms contribute to nonreciprocal charge transport.[23,50] The relation for a longitudinal component is described as $\sigma_{\parallel}^{(2)} = \eta_0 + \eta_1(\sigma_{\parallel}) + \eta_2(\sigma_{\parallel})^2$, where the coefficient $\eta_i$ denotes the respective part of the nonlinear conductivity.[50] First, we confirm that $\sigma$ is governed by the scattering time ($\sigma \propto \tau$) in the temperature range between 2 K and 8 K, as carrier concentrations is almost constant (**Figure 4a**). This allows the direct analysis on the transport mechanism regarding $\tau$. The $\tau^1$ contribution in the second term of the relation is attributed to the Berry curvature dipole, but it does not flip the sign of nonreciprocal response depending on the orientations of *N* nor *M*.[23,24] Thus, the Berry curvature dipole can be excluded from the origin of anomalous NLMC. The remaining first and third terms leave possible contributions from both the $\tau$-independent Berry connection polarizability, as well as from the $\tau^2$-dependent nonlinear Drude weight and anomalous skew scattering, respectively.[9,51,52] Figure 4b shows the plot for the scaling analysis fitted to the simplified relation $\sigma_\parallel^{(2)} = \eta_0 + \eta_2(\sigma_\parallel)^2$, in which the nonlinear longitudinal conductivity is defined as $\sigma_\parallel^{(2)} = (V_\parallel^{(2)} L)/(I_\parallel^2 R_\parallel^3)$, where *L* is the length of the channel.[23] Both the AFM (blue) and FM (red) states of the CrSBr(2L)/hBN display a non-zero intercept $\eta_0$, pointing to the presence of the Berry connection polarizability. On the other hand, a comparably negligible contribution from $\eta_2$ can be inferred from the slope of the line. Additional extrinsic contributions with multiple $\tau$ scaling have been recently proposed,[50,52] but the absence of tau-dependent contributions in the fitting suggest that such extrinsic contributions are negligible in our experiments. Therefore, and following this analysis, the origin of the anomalous nonreciprocal response in our CrSBr/hBN systems is attributed to the $\tau$-independent Berry connection polarizability,[25,31-33] which is often described as quantum metric dipole.[23-25]

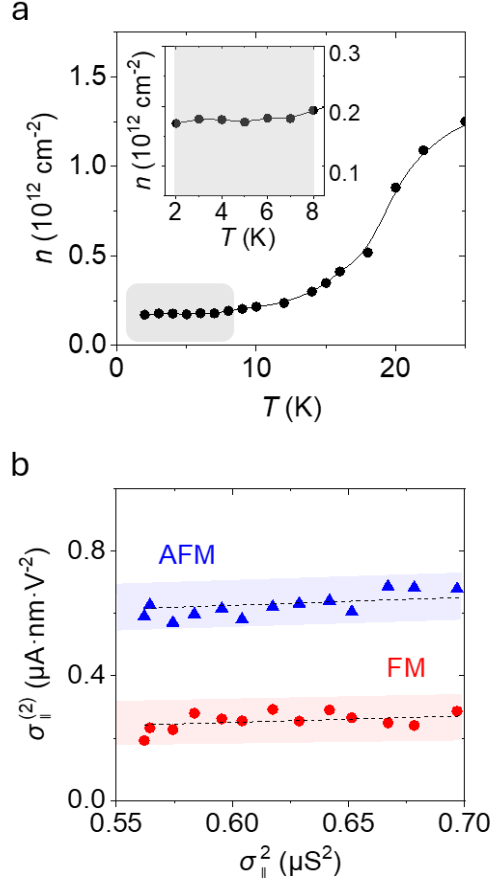

**Figure 4**. Nonlinear longitudinal conductivity from the Berry connection polarizability in CrSBr(2L)/hBN. a) Carrier concentration of CrSBr as a function of temperature. In a temperature range between 2 K and 8 K (grey area), carrier concentration is almost constant, indicating the scattering time determines the longitudinal conductivity. The inset is a magnified image for temperatures between 2 K and 8 K. b) Scaling analysis between the longitudinal nonlinear ($\sigma_{\parallel}^{(2)}$) and linear ($\sigma_{\parallel}$) conductivities for the antiferromagnetic (AFM) and ferromagnetic (FM) states of the CrSBr bilayer. The dotted lines are fitted lines from the scaling formula, $\sigma_{\parallel}^{(2)} = \eta_0 + \eta_2 (\sigma_{\parallel})^2$. The dominant contribution of Berry connection polarizability is observed as $\eta_0$, the origin intercept.

## 3. Conclusion

Our experiments present anomalous NLMC magnetic systems tailored on-demand. Thanks to the artificial symmetry breaking created by an interface with h-BN, we can reach a switchable second-order resistance in centrosymmetric ferromagnetic monolayer CrSBr. Additionally, we also reach four anomalous NLMC states in bilayer CrSBr by controlling both magnetization and Néel vectors in a single system, an unfulfilled challenge until now. The output response in the FM (AFM) state is three (one) orders of magnitude larger than previous reports in MnBi$_2$Te$_4$. This anomalous NLMC brings the use of nonlinear effects one step closer to commercial devices. Possible applications include reconfigurable radio frequency rectification, in which the collected direct current changes sign with the magnetic state of the rectifier. We also envision anomalous NLMC useful for the direct reading of the information encoded by complex magnetic materials. In particular, our development is suitable for AFM materials, which are actively sought-after for memory technologies due to their magnetic field immunity but whose information reading is extremely complex with standard techniques due to their zero net magnetization.

## 4. Experimental Section/Methods

*Device fabrication*: CrSBr and hBN flakes were mechanically exfoliated on cleaned Si/SiO$_2$ substrates in the argon-filled glovebox. To obtain a target thickness of a flake, we first confirmed the thickness of a flake using atomic force measurement and optical microscope. During the device fabrication, we used the optical contrast in the glovebox to avoid air exposure. Target flakes on the Si/SiO$_2$ substrate were picked-up using a polymer stamper consisting of a polycarbonate and a polydimethylsiloxane. A complete heterostructure was dropped down on top of a prepared substrate which had prepatterned electrodes. The electrodes consisted of Ti(2 nm)/Au(10 nm), fabricated using electron beam lithography and metal evaporation. The channel length of CrSBr was 2 μm and the width was between 1.5 μm and 2 μm for all devices. The main results in the text were presented from the devices; Device 1 for CrSBr(1L)/hBN in Figure 2, Device 11 for CrSBr(2L)/hBN in Figure 3 and Figure 4. Additional results from other devices are summarized in Table S2.

*Electrical measurement*: A fabricated device was mounted on the sample stage in a horizontal rotator of the physical property measurement system (Quantum Design). For direct current measurement, the Keithley 6221 was used to apply a current and the Keithley 2182 nanovoltmeter detected a voltage. Resistances with positive and negative direct currents were measured, named $R(+I)$ and $R(-I)$ respectively, and their average and a half of difference were denoted as $R^{(1)}$ and $R^{(2)}$, respectively. For alternating current measurement, the Keithley 6221 was used for current source, and a voltage was detected through the dual channel NF LI5660 lock-in-amplifier. The $R^{(2)}$ in the direct current measurement is equivalent to the $R^{2\omega}$ in the alternating current measurement[22] as $R^{(2)} = 2\sqrt{2}R^{2\omega}$ as $R^{2\omega} = V^{2\omega}/i^{RMS}$ where $i = \sqrt{2}i^{RMS}\sin\omega t$. The frequency was set to $\omega$ = 17 Hz.

**Acknowledgements**

This work was supported under Projects PID2021-122511OB-I00 and PID2021-128004NB-C21 funded by MICIU/AEI/10.13039/501100011033 and ERDF/EU.J.J. acknowledges the funding from the Ayuda FJC2020-042842-I funded by MCIN/AEI/10.13039/501100011033 and European Union NextGenerationEU/PRTR. F.J. is supported by Grant PID2021-128760NB0-I00 from the Spanish MCIN/AEI/10.13039/501100011033/FEDER, EU. M. S.-R. acknowledges support from La Caixa Foundation (No. 100010434) with code LCF/BQ/DR21/11880030. I.S. is supported by Grant No. PID2021-129035NB-I00 funded by MCIN/AEI/10.13039/501100011033. This work was also supported by the FLAG-ERA grant MULTISPIN, via the Spanish MCIN/AEI with grant number PCI2021-122038-2A, and by the Diputación Foral de Gipuzkoa (QUANTUM, project no. 2024-QUAN-000014-01). M. G. acknowledges support from the "Ramon y Cajal" Programme by the Spanish MCIN/AEI (grant no. RYC2021-031705-I). The authors acknowledge Elizabeth Goiri for helpful discussion.

Supporting Information

# Anomalous Nonlinear Magnetoconductivity in van der Waals Magnet CrSBr

*Junhyeon Jo\*, Manuel Suárez-Rodríguez, Samuel Mañas-Valero, Eugenio Coronado, Ivo Souza, Fernando de Juan, Fèlix Casanova, Marco Gobbi\*, and Luis E. Hueso\**

**Note S1**. **Symmetry analysis of NLMC and anomalous NLMC.**

The general relationship between electric field and current density in the presence of magnetic field is written as $E_i = \rho_{ij}(\vec{J}, \vec{B}) J_i$ where $\rho_{ij}$ is the resistivity and $i, j = x, y, z$ represent cartesian directions in the crystal coordinate system, and repeated indices are summed over. In an orthorhombic system like CrSBr where the three lattice vectors are mutually perpendicular one may also take $i = a, b, c$. The symmetric part of the resistivity $\rho_{ij} = \rho_{ji}$ can be expanded in powers of the current and magnetic field as

$$\rho_{ij}(\vec{J}, \vec{B}) = \rho_{ij}^{(1)} + \rho_{ijkl}^{(1),B} B_k B_l + \rho_{ijk}^{(2)} J_k + \rho_{ijkl}^{(2),B} J_k B_l$$

where the second term is the usual magnetoresistance, the third is the zero field nonlinear resistance, and the fourth represents NLMC. These last two terms are only allowed when inversion symmetry is broken.

The crystal structure of bulk CrSBr has orthorhombic space group $P_{mmn}$, with point group $D_{2h}$, and a free-standing monolayer has the same symmetry. The presence of the substrate breaks inversion and reduces the symmetry to $C_{2v}$, with two-fold rotations around the normal $C_{2c}$ as well as mirror planes $M_a$ and $M_b$. For in-plane transport along the $a$ direction, the only component that is allowed by symmetry is $\rho_{aaab}^{(2),B}$ while $\rho_{ijk}^{(2)} = 0$ due to $C_{2c}$. When the magnetic field is along $b$ then only $J_a$ and $B_b$ are non-zero and we can write $\rho_{aa}(\vec{J}, \vec{B}) = \rho_{aa}^{(1)} + \rho_{aabb}^{(1),B} B_b B_b + \rho_{aaab}^{(2),B} J_a B_b$ (note $a, b$ represent fixed directions and are not indices to sum over). Since this is the only geometry we probe in this work, we simplify this equation to

$$\rho = \rho_0(1 + \alpha B^2 + \gamma' BJ)$$

where $\rho_0 = \rho_{aa}^{(1)}$, $\alpha = \rho_{aabb}^{(1),B}/\rho_{aa}^{(1)}$ and $\gamma' = \rho_{aaab}^{(2),B}/\rho_{aa}^{(1)}$.

Finally, in a geometry where current flows through a cross-section A and the voltage is measured between contacts separated by distance L, the total current is $I = AJ$ and the voltage $V = LE$ so the resistance defined as $V = RI$ is $R = \rho L/A$. With this can finally write

$$R = R_0(1 + \alpha B^2 + \gamma IB)$$

where $\gamma = \gamma'/A$.

In the ferromagnetic state of the monolayer, we can similarly expand $\rho_{ij}(\vec{J},\vec{B},\vec{M})$ to first order in magnetization $\vec{M}$, and since $\vec{B}$ and $\vec{M}$ have the same symmetry, we arrive to an analogous equation

$$R=R_0(1+\alpha_M M^2+\gamma_M IM)$$

where $\gamma_M$ represents anomalous NLMC. Note in the presence of both magnetization and magnetic field the equation would contain all terms from both expansions, as well a mixed term $\alpha_{MB}MB$.

Finally, in the case of the AFM bilayer the Néel vector $\vec{N}$ points in the $b$ direction and has the same symmetry as $\vec{M}$ under $C_{2v}$, so the same equation can be written for the Néel vector

$$R=R_0(1+\alpha_N N^2+\gamma_N IN)$$

where $\gamma_N$ represents anomalous NLMC. This equation would still be valid even in the absence of a substrate with group $D_{2h}$, because $\vec{N}$ is odd under inversion.

On the other hand, for the FM state in both the monolayer and bilayer, $\gamma_M=0$ without a substrate because $\vec{M}$ is even under inversion.

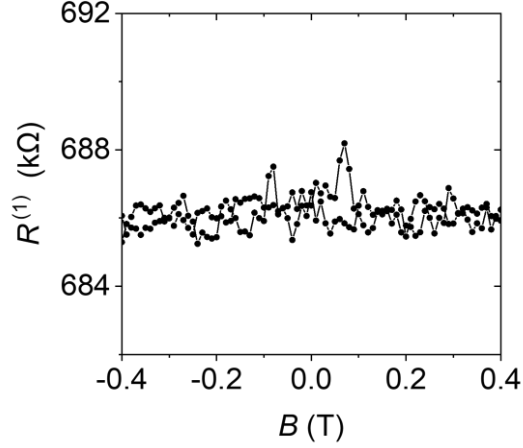

**Figure S1**. Average resistance in CrSBr(1L)/hBN. The average resistance is defined as $R^{(1)} = [R(+I) + R(−I)]/2$, based on $R(+I)$ and $R(−I)$ data collected from Figure 2b.

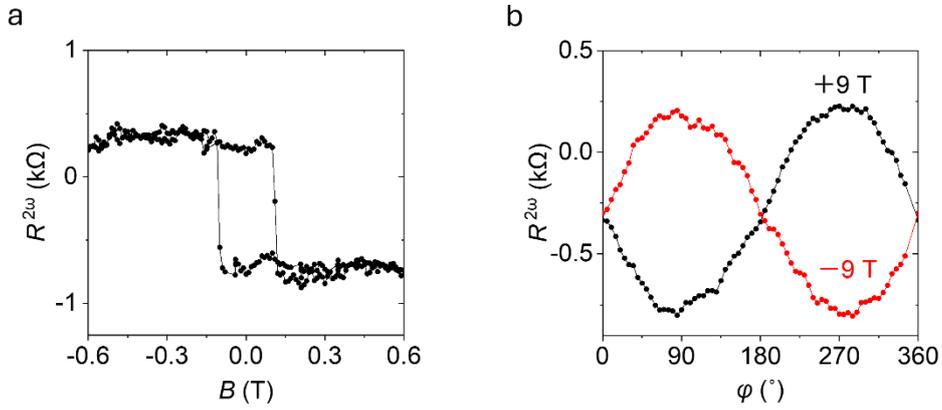

**Figure S2**. Second harmonic resistance measurement in CrSBr(1L)/hBN. a) Second harmonic resistance ($R^{2\omega}$) as a function of a magnetic field ($B$). b) $R^{2\omega}$ as a function of an in-plane angle ($\varphi$). The second harmonic measurement displays equivalent results to the measurement with direct currents, shown in Figure 2c and Figure S4. The $\varphi$ is defined in Figure 2a.

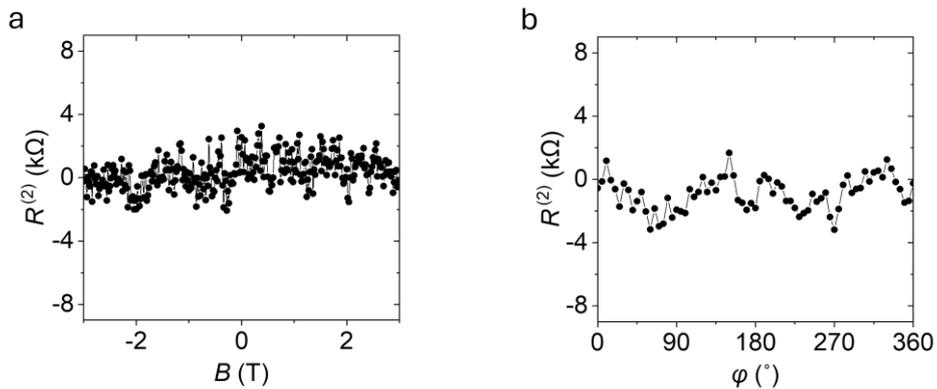

**Figure S3**. Anomalous nonlinear magnetoconductivity measurement of CrSBr(1L) without an hBN layer. a) Resistance difference, $R^{(2)} = [R(+I) − R(−I)]/2$, as a function of a magnetic field ($B$). There is no feature of hysteresis. b) $R^{(2)}$ as a function of an in-plane angle ($\varphi$). There is no detectable response. The absence of $R^{(2)}$ implies that inversion symmetry is preserved in the pristine CrSBr(1L). The $\varphi$ is defined in Figure 2a.

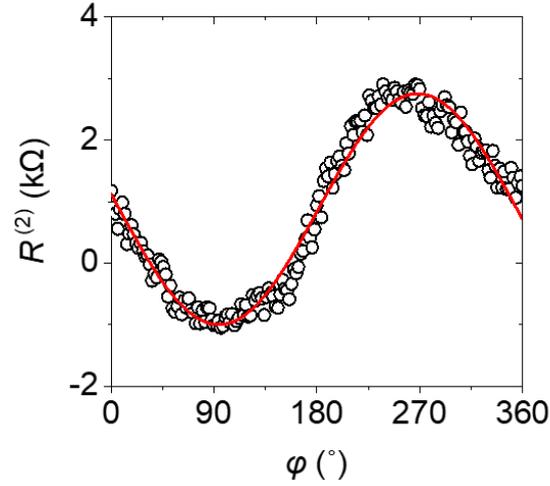

**Figure S4**. Angular dependence nonlinear magnetoconductivity in CrSBr(1L)/hBN. Angle ($\varphi$)-dependent second-order $R^{(2)}$ under an in-plane magnetic field of 9 T. The applied field of 9 T is higher than the in-plane coercive fields (for both *a*- and *b*-axis) of CrSBr. When the magnetization is aligned perpendicular to the current direction ($\varphi = 90°$ and $270°$), the $R^{(2)}$ reaches its maximum value. On the other hand, when the magnetization is aligned along the crystal *a*-axes ($\varphi = 0°$ and $180°$), the $R^{(2)}$ response vanishes. The $\varphi$ is defined in Figure 2a.

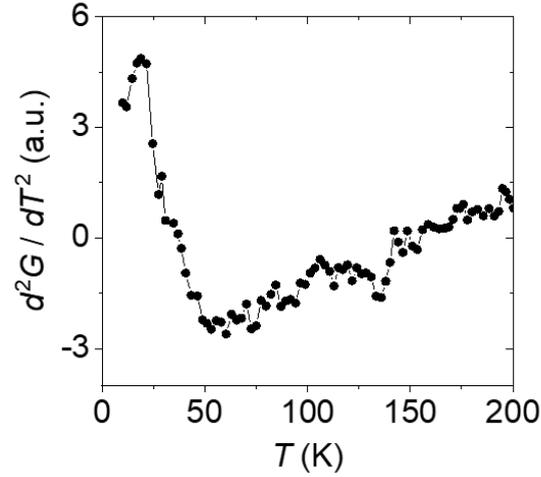

**Figure S5**. Second derivative of the conductance of CrSBr(1L)/hBN. The curve shows zero crossing and a dip-like feature close to 140 K which is close to the reported Curie temperature of CrSBr(1L).[27] We assume that the difference between the Curie temperature and the disappearance temperature of $R^{(2)}$ is caused by the $R^{(2)}$ signal being too small to be detected accurately.

| Magnetic state | Type | Material | T (K) | B (T) | $R^{(2)}$ (Ω) | $R^{(2)}/j$ (Ωm²/ A) | $\gamma'M$ (A⁻¹·m²) | $T_{cri}$ (K) | Reference |
|---|---|---|---|---|---|---|---|---|---|
| Ferromagnetic | Magnetic topological insulator (edge state) | Cr$_x$(Bi,Sb)$_{2-x}$Te$_3$ | 2 | 0 | 4,100 | 3.28E-09 | 2.05E-09 | 40 | [22] |
| | | MnBi$_2$Te$_4$(5L)/hBN | 11 | 0 | 200 | 3.50E-10 | 1.75E-11 | 23 | [30] |
| | Layered magnet | CrSBr(1L)/hBN | 2 | 0 | 3,390 | 4.06E-06 | 5.93E-12 | 100 | This work |
| | | CrSBr(2L)/hBN | | 0.3 | 1,930 | 6.80E-06 ★ | 1.90E-11 | 140 | This work |

| Magnetic state | Type | Material | T (K) | B (T) | $R^{(2)}$ (Ω) | $R^{(2)}/j$ (Ωm²/ A) | $\gamma'N$ (A⁻¹·m²) | $T_{cri}$ (K) | Reference |
|---|---|---|---|---|---|---|---|---|---|
| Antiferromagnetic | Magnetic topological insulator | MnBi$_2$Te$_4$(4L)/hBN | 2 | 0 | 120 | 5.38E-07 | 4.80E-11 | 23 | [23] |
| | | MnBi$_2$Te$_4$(6L)/hBN | 1.6 | 0 | 90 | 1.58E-06 | 1.06E-09 | 21 | [25] |
| | Layered magnet | CrSBr(2L)/hBN | 2 | 0 | 4,390 | 1.41E-05 ★ | 4.12E-11 | 140 | This work |

**Table S1**. Comparison of anomalous nonlinear magnetoconductivity in magnetic systems. Second-order resistance is defined as $R^{(2)} = [R(+I) - R(-I)]/2$. The $j$ is the current density, and the $A$ is the cross-sectional area of a sample, defined as $A = tW$ where $t$ is thickness and $W$ is the width of a conducting channel. The width of the edge state of topological insulator was considered as 10 nm. The normalized coefficients are described as $\gamma'_M M = R^{(2)}/(R_0 j)$ for the FM state and $\gamma'_N N = R^{(2)}/(R_0 j)$ for the AFM state. The star marks indicate the highest values in each magnetic state.

| Magnetic state | Structure | Device | B (T) | $R^{(2)}$ (Ω) | $R^{(2)}/j$ (Ωm²/ A) | $\gamma'M$ (A⁻¹·m²) |
|---|---|---|---|---|---|---|
| Ferromagnetic | CrSBr(1L)/hBN | Device 1 | 0 | 3,390 | 4.06E-06 | 5.93E-12 |
| | | Device 2 | 0 | 1,880 | 2.41E-06 | 1.19E-11 |
| | CrSBr(2L)/hBN | Device 11 | 0.3 | 1,930 | 6.80E-06 | 1.90E-11 |
| | | Device 12 | 0.3 | 1,980 | 4.43E-06 | 2.28E-11 |
| | | Device 13 | 0.3 | 3,650 | 8.56E-06 | 1.98E-11 |

| Magnetic state | Structure | Device | B (T) | $R^{(2)}$ (Ω) | $R^{(2)}/j$ (Ωm²/ A) | $\gamma'N$ (A⁻¹·m²) |
|---|---|---|---|---|---|---|
| Antiferromagnetic | CrSBr(2L)/hBN | Device 11 | 0 | 4,390 | 1.41E-05 | 4.12E-11 |
| | | Device 12 | 0 | 3,760 | 8.42E-06 | 4.20E-11 |
| | | Device 13 | 0 | 5,830 | 1.37E-05 | 3.00E-12 |

**Table S2**. | Summary of anomalous nonlinear magnetoconductivity of multiple devices. The values were collected from the data at 2 K, and reproducible results are displayed for each heterostructure.